\documentclass[aps,superscriptaddress,prl,twocolumn]{revtex4-1}

\usepackage{amsbsy}
\usepackage{amsfonts}
\usepackage{amssymb}
\usepackage{amsmath}
\usepackage{color}
\usepackage{graphicx}
\usepackage{xspace}
\usepackage[normalem]{ulem}

\usepackage{dsfont}

\usepackage{numprint} 

\newcommand{\bb}[0]{\begin{eqnarray}}
\newcommand{\ee}[0]{\end{eqnarray}}

\begin{document}

\title{What is quantum in quantum randomness?} 

\author{Philippe Grangier}
\affiliation{Laboratoire Charles Fabry, IOGS, CNRS, Universit\'e Paris Saclay, F91127 Palaiseau, France}

\author{Alexia Auff\`eves}
\affiliation{CNRS and Université Grenoble Alpes, Institut Néel, F-38042 Grenoble, France}
\email{alexia.auffeves@neel.cnrs.fr}

\begin{abstract}
It is often said that quantum and classical randomness
are of different nature, the former being ontological
and the latter epistemological. However, so far
the question of "What is quantum in quantum
randomness", i.e. what is the impact of quantization and discreteness 
on the nature of randomness, remains to answer. In a
first part, we explicit the differences between quantum
and classical randomness within a recently proposed ontology
for quantum mechanics based on contextual objectivity. In this view, quantum
randomness is the result of contextuality and
quantization. We show that this approach strongly impacts the purposes of quantum theory as well as its areas of application. In particular, 
it challenges current programs inspired by classical reductionism, aiming at 
the emergence of the classical world from a large number of quantum systems.
In a second part, we analyze quantum physics and thermodynamics as theories of randomness, unveiling their mutual influences.
We finally consider new technological applications of quantum randomness opened in the emerging field of quantum thermodynamics.
\end{abstract}

\maketitle

\section{Introduction}

When something unpredictable happens to us, we can
invoke fate, or randomness. Fate has some advantages: It
allows not blaming ourselves for not taking the umbrella
while walking under heavy rain, or for playing the
red when the black wins. On the contrary, randomness
leaves open the possibility that we could have made a
different choice, if we had known - and then we would
have remained dry, and become rich. In this view, the
manifestations of randomness are subjective, for they
depend on our information on the course of events:
Ultimately, complete information erases randomness.
In everyday life, as well as in the classical realm, randomness and information thus appear
as two antinomic concepts, randomness being accidental and due to an incomplete knowledge of things. Such "epistemic randomness" lies at the root of statistical physics,
where randomness is due to the irreducible loss of information taking place while zooming out from the microscopic
to the macroscopic scale. This motivates the tenants of reductionism to make use of this methodology as a means to reach the ultimate level of physical reality, where randomness vanishes and the laws of physics become reversible. \\

\noindent \textbf{Early views on quantum randomness}
Nowadays it is usually advocated that quantum and classical
randomness are of different nature. However in the early times of quantum mechanics, quantum randomness was presented as the irreducible 
disturbance induced on the quantum state by the act of measuring it. The most paradigmatic example of this is the famous Gedanken experiment
of Heisenberg's microscope. 
But invoking such measurement induced disturbance was
actually making quantum randomness very similar to classical randomness. In both cases, the
"real" states are hidden at the microscopic level, and the information on these states is irreversibly lost when reaching the macroscopic world. In this view, quantum measurement
appeared as some sort of coarse graining, calling for the search of hidden variables and the attempt to get rid of what was called the "measurement problem". \cite{EPR}.\\

\noindent Since then, local hidden variable models have been discarded through the multiple successful
experimental violations of Bell's inequalities \cite{Bell, AGR81,AGR82,ADR82,Hanson, Zeilinger, Nam}, and for many quantum physicists it is now well established that the world is non local. In such a non-local world, quantum randomness has a clear justification, for it prevents parties from communicating faster than light and protects Einstein's causality \cite{Gisin}.
Actually, the fact that nature is intrinsically random is taken for granted in various reconstructions of quantum mechanics, which is envisioned within the framework of
generalized probability theories \cite{Hardy, Rovelli}. However, the causes of randomness in the quantum world - as well as the very impact of quantization and discreteness on the nature of quantum randomness, are rarely discussed. Actually no alternative, physical explanation has been proposed that would replace the Heisenberg's microscope.\\

\noindent Just like non-locality, the idea that the quantum world is contextual spread 
since Kochen-Specker theorem \cite{KS}. 
{\bf It is the purpose of the present paper to explain randomness in such a contextual world}. In particular, we aim at unveiling the causes of quantum randomness and investigate its essential differences with classical randomness. 
Answering these questions is challenging. After all, apart from fundamental indeterminism (that appears for instance in the antic example of the clinamen), randomness is most often attributed to some ignorance on the state of a system: Our deepest intuitions on randomness are thus built in a non-contextual world.\\

\noindent We shall base the discussion on the recently developed "Contexts-Systems-Modalities (CSM)" approach \cite{CSM1,CSM2,CSM3}, which proposes a new ontology for
quantum mechanics based on a new perspective on objectivity, called "contextual objectivity". \cite{CO}. In the present paper, we show that quantum randomness is the natural byproduct of
both the contextual nature of states, and the quantized nature of elementary objects, unambiguously answering the question: "What is quantum in quantum randomness?". 
Our approach challenges the reductionist idea that the
classical world should emerge from a large collection of quantum systems, whose evolution would be intrinsically governed by the Schr\"odinger equation. 
In a second step, we compare the treatment of randomness in thermodynamics and quantum physics.  From a technological perspective, a usually mentioned application of quantum randomness is provided by random number generators. Here we suggest new potential applications by considering recent proposals in quantum thermodynamics, where quantum noise appears as an energetic resource fueling new kinds of quantum engines.

\section{Contextual Objectivity}
In this section we set up the stage and define the notion of contextual objectivity, based on a primitive ontology of objects and states. 

\subsection*{Objects and Phenomena}
Our most basic assumption is the existence of a natural world. This world is objective, i.e. it exists independently of any observer. 
The world can be analyzed in systems, i.e. finite entities which are also objective. Since it is a finite entity, a system is surrounded by a large number of other systems that we call a context. Note that as a finite entity, a context can also play the role of a system, and be embedded in another context.\\

\noindent On the other hand, the natural world is the theater of phenomena, i.e. actual physical events. Since they are actual and objective, phenomena are unique, well-determined, and they exist independently of any observer.  A system and its context define a possible partition of the world, which may be considered as objective (and not fictitious) when it gives rise to such well defined phenomena.\\

\subsection*{States}
Let us consider such an objective partition consisting of a system and a context. This partition is quite generic and can describe situations of the everyday life: For instance, a table as the system, and as possible contexts, two cameras measuring "Position" and "Color". It can also capture the situations encountered in the lab, involving a quantum system (like an electronic spin) while the context is the measuring apparatus (like a Stern-Gerlach apparatus). To grasp the meaning of a state, we adopt an operational approach and suppose we can act on the settings of the context, giving rise to different phenomena. A given setting at time $t_i$ is denoted $C(t_i)$, and corresponds to a "question" (For instance, which camera is on determines if the question is "Color?" or "Position?", the angle of the SG fixes in which basis the spin is measured..). The resulting phenomenon $X_{C(t_i)}(t_i)$ is the corresponding "answer". The complete sequence of collected answers $[X_{C(t_i)}(t_i)]$ is further called an "ID card".\\

\noindent We now suppose that an identical question repeatably gives rise to the same identical answer (e.g., "Color?" is systematically "White"). Such repeatability is usually explained by postulating the existence of some objective {\bf state} ("The table is white"). It signals an "element of reality" \cite{EPR}, allowing to predict with certainty the answer ("White") to the same question ("Color?"). This ontological concept of state "upgrades" the operational concept of ID card.

\subsubsection*{Classical phenomenology}
In the classical phenomenology we are accustomed to, the same question triggers the same answer, independently of the ordering of the questions. In other words: \\

\noindent {\bf Condition 1: Classically, if $C(t_i) = C(t_j)=C_i$ then $X_{C_i}(t_i) = X_{C_i}(t_j)=X_i$, whatever $C(t_k)$ where $t_i < t_k < t_j$.}\\

\noindent In this situation, the ID card can be upgraded into a very special kind of state $[X_i]_{\{C_i\}}$, gathering all answers to all questions and corresponding to embedding the system in all contexts $\{C_i\}$ simultaneously. For instance, "White and in the back of the room" is the state revealed by the context corresponding to asking "Color?" and "Position?" simultaneously. In this view, there is no need to keep track of the context (i.e., which camera is on) to define the state, which is therefore seen as pertaining to the system alone. 

\noindent The non-contextuality of states characterizes classical ontology. From our reasoning, it should nevertheless appear that it selects a very special class of eligible states. It is solely because we perceive classical phenomena continuously, and for free, that classical ontology has historically won over other ontologies and lies at the root of our deepest intuitions on the nature of physical reality.  In what follows, we extend our ontology of states beyond such classical intuitions. For now, we use the notion of non-contextuality to add precision to our definition of context:\\

\noindent \textbf{Postulate 0. Contexts}\\
{\it By definition, contexts are characterized by non-contextual states}.\\
The Postulate 0 allows allows a more precise characterization of the protocol used above to build an ID card and define states. As a matter of fact, before being used as a context, this entity must have primarily played the role of a system, and been systematically tested by other contexts. The corresponding ID card must fulfill the Condition 1. Eventually, the probed entity used as a context is characterized by its own state, which does neither depend on some other context it would be embedded in, nor on the system it may possibly contain. Postulating the non-contextuality of the context allows breaking the recursive character of our construction, providing the necessary fixed point to build a general ontology of states. It also eliminates {\it per se} the paradox of Wigner's friend and related problems, since the state of a context defined that way is objective, and consequently, identical for any observer. 

\subsubsection*{Quantum phenomenology}
Now that contexts have a well defined state, it is possible to provide a more precise description of the situation under study: A system is embedded in a context previously probed as mentioned above. The state of the context changes in time, defining a sequence $C(t_i)$ and a corresponding sequence of phenomena $X_{C(t_i)}(t_i)$. In the usual quantum phenomenology, it is well known that the registered phenomena depend on the ordering of the context's state, such that Condition 1 is not valid anymore. For instance, two successive measurements of the x-component of a spin will provide identical results, but this will not necessarily be the case if a measurement of the y-component is interposed. There are at least three options to acknowledge for this non-classical behavior:

\begin{enumerate}
\item {\bf Na\"ive realism}: The system has a state, the context perturbs this state. This view leads to the search for hidden variables, and the desire to get rid of the context.
\item {\bf Instrumentalism}: There are no states, only correlations between preparations and measurements. In this view, the ID card is never upgraded and physics operates without any ontological basis.
\item {\bf Contextual objectivity} goes beyond the observation of correlations and acknowledges it is possible to obtain repeatable phenomena, provided the context is not changed: This is the textbook situation of quantum non-demolition (QND) measurements. In this case, the ID card can be upgraded into a state, but only for a given setting/state of the context. This leads us to our second postulate:
\end{enumerate}

\noindent \textbf{Postulate 1. Systems, Contexts, Modalities}\\
{\it In general, states label both a system and a context. To distinguish them from classical states which are non-contextual, we call a contextual state a modality. Since it is a phenomenon, the modality is as objective as a non-contextual state: This general ontology of states characterizes Contextual Objectivity \cite{CSM1,CO}. Because they are actual, modalities appearing within the same given context are mutually exclusive: If one is realized, the others are not. On the other hand, no conclusion can be drawn about modalities pertaining to two different contexts, because these two contexts cannot be simultaneously realized.}\\

\noindent We have defined the contextual and non-contextual ontologies of states, building on the quantum and the classical phenomenology respectively. However, note that these ontologies can be conceived in full generality, the classical and the quantum world corresponding to specific realizations. After characterizing the context, we now focus on "Elementary systems" to formulate the third and last postulate of our ontology of states:\\

\noindent \textbf{Postulate 2: Elementary systems} {\it We call "elementary" those systems that are characterized by a discrete, finite number N of exclusive modalities. This number is independent of the context the system is embedded in. Note that this elementary character is given through their phenomenology and does not condition the "stuff" they are made of.}\\

\noindent These three postulates capture a critical, though textbook situation encountered in the quantum world: An elementary system embedded in a context, giving rise to a discrete and finite amount of modalities. We now demonstrate that this situation carries all the seeds for new kinds of randomness: Ontological randomness and quantum randomness, which provide the natural soil of the quantum formalism.

\begin{figure}[htb]
    \includegraphics[scale=0.3]{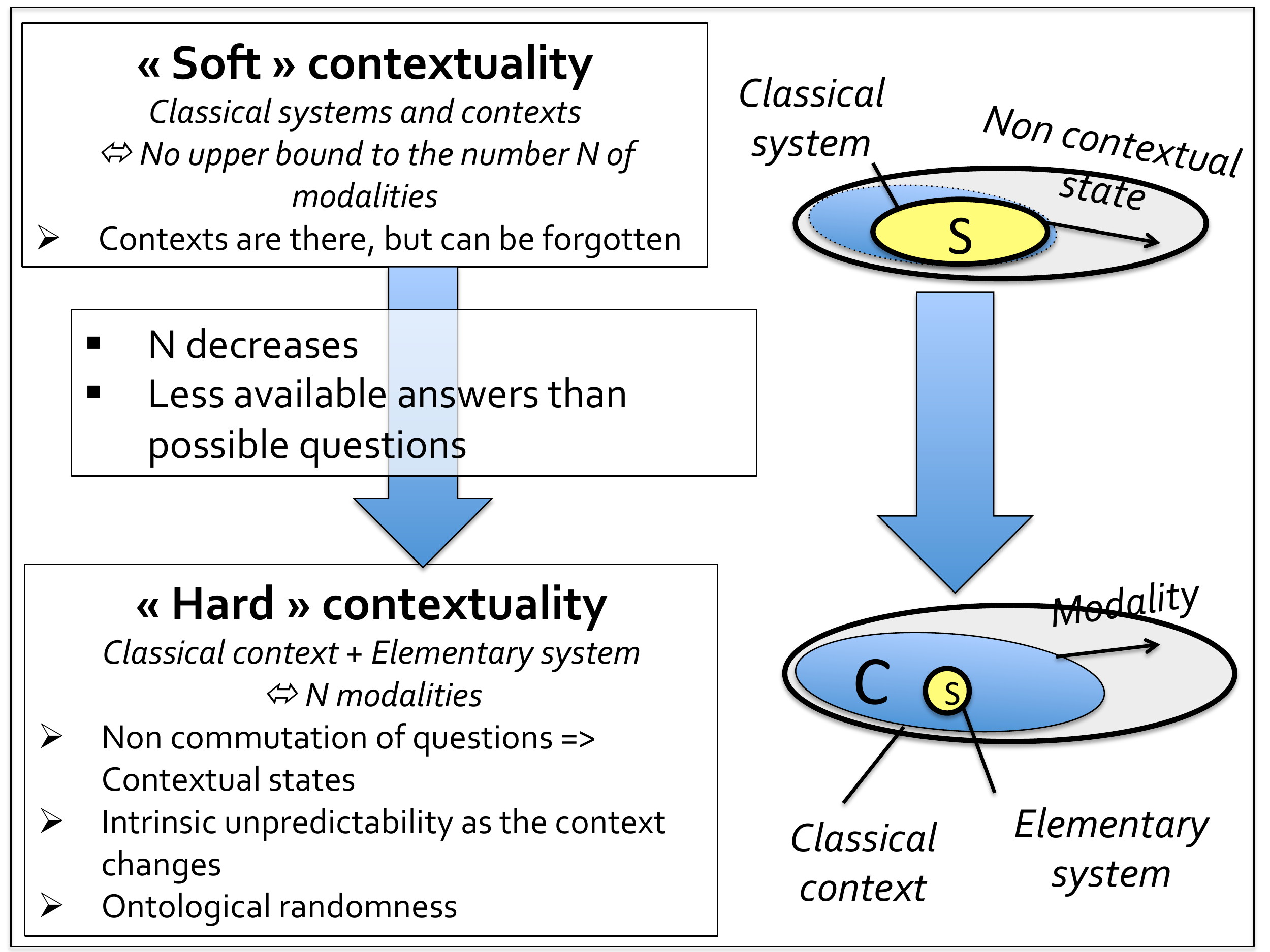} 
    \caption{\label{fig_universal}
       Soft vs. hard contextuality. Contextuality is a universal feature of our ontology of states, all states being revealed within a given context. Soft and hard contextuality correspond to the classical and the quantum phenomenology respectively. Quantum phenomenology is observed when the number of available modalities is reduced down to the point where the system becomes elementary. }
\end{figure}

\section{Randomness in a contextual world}

\subsubsection*{Ontological randomness}
In the previous section, we have pinpointed two special types of states: contextual vs. non-contextual. In the former, a reference to the context is needed to define a state: We shall talk about "Hard contextuality", which is typical of quantum phenomenology. In the latter, contexts are always here, but can be forgotten, which corresponds to classical phenomenology: We shall talk about "Soft contextuality". It thus appears that there is no intrinsic difference of nature between the two classes of states, contextuality being universal: In other words, a context is always already there. We now demonstrate that the appearance of hard contextuality is actually controlled by the interplay between the number of the system's modalities and the number of the context's potential states (Fig.1).\\

\noindent \textbf{Proof by contradiction} We consider a system characterized by $N=2$ for the sake of simplicity, embedded in a context whose state can be changed from $C_1$ to $C_2$. Modalities in $C_i$ with $i=1,2$ are denoted $X_{C_i}$. We choose $X=Yes, No$.  At time $t_1$ (resp. $t_2$) the system embedded in $C_1$ (resp. $C_2$) gives rise to $X_{C_1}(t_1)$ (resp. $X_{C_2}(t_2)$).  At time $t_3$ the context is prepared back in the state $C_1$. The question is: Can we predict with certainty which modality will be realized?

Let us suppose we can. In this case, there would actually be $4$ exclusive modalities corresponding to the joint events $\{ X_{C_1},X_{C_2} \}$. This violates our second postulate, which puts an upper bound to the amount of exclusive modalities for elementary systems. Therefore, the modality realized when the context's state is changed from $C_2$ to $C_1$ cannot be predicted with certainty.
The quantization of the modalities for an elementary system thus induces the non-commutation of the questions (otherwise, the answer would be $X_{C_1}(t_1) = X_{C_1}$ with certainty while coming back to $C_1$). This non-commutation is characteristic of hard contextuality. \footnote{Note that in this simple example involving $2$ modalities, it is possible to mimic ontological randomness by epistemic randomness in the spirit of Spekkens' toy model \cite{Spekkens}. Following Spekkens' terminology, the system would be characterized by 4 "ontological states" $\{ X_{C_1},X_{C_2} \}$, but the measurement only gives access to two "epistemic states" (either $X_{C_1}$, or $X_{C_2}$ ), and measurement randomly perturbs the state as the context changes. As mentioned above, the same kind of event (random change of phenomenon as the context changes) can give rise to different interpretations, this one being reminiscent in some sense of our "na\"ive realist" approach (The system has a state, and measurement perturbs it). Our attempt is to deconstruct such intuitions inspired by the classical phenomenology. }\\

\noindent This simple reasoning has deep consequences. It basically states that in a contextual world, quantizing the amount of exclusive modalities of a system gives rise to some intrinsic unpredictability signaling {\bf ontological randomness}. While intrinsic randomness is postulated from metaphysical consideration, while epistemic randomness is caused by the ignorance of some hidden state, the cause of ontological randomness can be caught by the following simple sentence: {\bf There are less available answers than possible questions}. More precisely, the number of possible answers to all possible questions is larger than the number of allowed mutually exclusive answers for the considered system. As a consequence, some of the answers are not mutually exclusive, and thus must be related in a probabilistic way - this probability reflecting by no means a lack of information.\\

\noindent To summarize, we now draw the prescriptions to observe ontological randomness: 
\begin{enumerate}
\item An elementary system characterized by $N$ exclusive modalities is embedded in a context.
\item The state of the context changes from $C_1$ to $C_2$.
\end{enumerate}

\noindent In the primitive scenery described by the conditions (i) and (ii), there is only one kind of physical event to describe, i.e. the random change of modality when the context changes. The ultimate goal of the corresponding "primitive theory" is to provide a mathematical description of this primordial random event, i.e. an $N$ by $N$ probability matrix gathering the conditional probabilities of jumping from one modality $X_{C_1}$ in the initial context to another modality $Y_{C_2}$ in the subsequent context.

\subsubsection*{Quantum randomness and quantum theory}
Ultimately, {\bf quantum randomness} is a special case of ontological randomness, where the probabilities at the change of context follow Born's rule. There is only one condition to add to (i) and (ii) to single out quantum randomness and recover the quantum formalism: \\

(iii) The state of the context can change {\bf continuously}\\

\noindent The conditions (i)-(iii) are realized in most experiments of quantum mechanics, which involve by construction an elementary system (e.g. an electronic spin) and an experimental context, whose state is non-contextual and can be changed continuously (e.g. a Stern-Gerlach apparatus, whose state is described by the angle of the magnetic field). When the condition (iii) is added, the "primitive theory" mentioned above is nothing but quantum theory. The mathematical demonstration is available in \cite{CSM2,CSM3} and will not be recalled here. It boils down to evidencing that the probability matrix is uni-stochastic, i.e. its expression involves the complex unitary matrices characterizing the quantum formalism. To show this result, the continuity argument is essential. \\  

\noindent Interestingly, it appears that while intrinsic and epistemic randomness grow in a non-contextual world, ontological and quantum randomness are natural byproducts of the quantized nature of contextual states, clearly answering the question "What is quantum in quantum randomness?". At this point, we emphasize that major usually postulated features of the quantum theory (probabilistic nature, non-commutation of the questions) have been derived here without invoking any mathematical formalism. On the contrary, they have been deduced from from phenomenological postulates (contextuality, quantization) inspired from canonical quantum experiments. Interestingly, this conceptual, non-mathematical approach leads to revert well-established hierarchies in the program of quantum theory:

\begin{itemize}
\item Deterministic, unitary evolutions are nothing but calculation tools involved in the mathematical description of this primordial event. This is an important change of perspective with respect to many interpretations of quantum mechanics such as many-worlds or quantum darwinism, that aim to eliminate stochastic and non-unitary transformations from the quantum postulates. It also goes against classical intuitions drawn from statistical physics, where determinism governs true events at the microscopic level, while randomness reflects the imperfection of our description. {\bf Within CSM, randomness and non-unitarity come first, determinism and unitarity second.}\\

\item In the same way, CSM reverts the hierarchy between quantization and interferences. In quantum mechanics, quantization is often presented as a consequence of the superposition principle and the confinement of physical systems. Interestingly here, complex numbers associated to interferences are a consequence of the quantization of exclusive modalities.  \textbf{Within CSM, quantization comes first, interferences second.}\\

\item Finally, we emphasize an essential difference with respect to quantum ontologies where the wave function represents a real property of a system, that is collapsed by the measurement. These ontologies save the non-contextuality of states; however, the real wave function is hidden, and describes a potentiality. Within CSM, the real is actual, and the theory is all about phenomena - the price to pay being the contextual nature of states. The wave function, just like unitary matrices, is a calculation tool which has no physical existence. There is nothing like wave packet reduction, but stochastic jumps from modality to modality. This makes CSM highly compatible with quantum trajectory based theories, such as quantum stochastic thermodynamics (See below). \textbf{Within CSM, actual comes first, potential second.}
\end{itemize}

\subsubsection*{Field of application of quantum theory}

At this point it is good to take a step aside and draw a first conclusion of our findings: We have shown that when coupling an elementary system to a context whose state can change continuously, a new kind of randomness appears: Quantum randomness, and that the ultimate goal of quantum theory is to provide a mathematical description of the random phenomena appearing in this scenery. Importantly, our approach fundamentally restricts the field of application of the quantum formalism, namely to partitions involving a system and a context.  \\

\noindent First of all, let us notice that the system/context partition is not restricted to the lab, but exists in nature: A photon propagating through a birefringent material, a electron in the potential field of a nucleus..are nothing but systems coupled to their respective environments. These environments play the role of natural contexts, measuring in their fixed "pointer basis" \cite{WZ1}, while modalities naturally evolve as time flies \cite{CSM1}. Therefore, phenomena described by the quantum formalism show up in the natural world without the intervention of any external agent. However, it does not reflect an absolute property of systems, but joint properties of systems and contexts. \\

\noindent Interestingly, the CSM approach sheds new light on the quantum-classical boundary. This powerful concept has been introduced to describe the transition from the quantum to the classical phenomenology, by considering larger and larger systems. In the so-called decoherence theory, the quantum to classical transition is attributed to the measurement performed by the environment on the system, which becomes unavoidable when the system is large \cite{WZ1}. Owing to its numerous experimental demonstrations \cite{LKB}, decoherence has for long fed the idea that the classical world- {\bf including the experimental context} could ultimately be rebuilt from larger and larger samples of systems. Such a bottom-up move is typical of statistical physics and reductionist theories  (See below).

In some sense, the CSM approach reverses this move, since a classical context is always already there: It all starts at the classical level. The context embeds a system, whose number of modalities is decreased, up to the point where the system is elementary and quantum effects such as hard contextuality appear. In this top-down move, ontological, quantum randomness, so as the whole quantum formalism are the consequences of the fundamental mismatch between the discrete nature of the exclusive modalities characterizing elementary systems, and the continuous nature of the states of a context. \\

\noindent In this view, the quantum formalism does not capture any absolute feature of the world. In particular, the Schr\"odinger equation does not describe the absolute evolution of a system, its {\it raison d'\^etre} being a change of context. There is no wave function of a system such as the whole Universe either. On the contrary, as part of the quantum formalism, the existence of a wave function is restricted by nature, at the interface between a context and a system. Note that this represents an important evolution with respect to standard Copenhaguian approaches, where the presence of the context is necessary for practical reasons. Within CSM, the context is necessary for fundamental reasons, for it conditions the appearance of ontological randomness.\\

\noindent This conceptual jump precludes any attempt to rebuild the classical world from elementary quantum systems, whose evolution would be governed by Schr\"odinger equation \cite{WZ2,PT}. As argued above, Schr\"odinger equation, {\it aka} unitary evolutions in quantum mechanics are solely involved in the modeling of the probability matrix describing a change of context, and have no absolute meaning. In this view, the search to dissolve the context in systems is meaningless. Even if the system and the context are ultimately made of the same "stuff", they do not share the same status in the ontology of states, the context being characterized by non-contextual states, and not the system. Note that this is an important difference with respect to the {\it Relational Quantum Mechanics} reconstruction \cite{Rovelli} which involves relations between equivalent systems, and postulates quantum randomness. On the contrary, CSM postulates the difference of status between system and context, to deduce quantum randomness from their coupling. Because of this primordial heterogeneity in the world's description, the CSM approach challenges classical reductionism, which naturally grows in a homogeneous, non-contextual world.\\

\section{Theories of randomness: thermodynamics vs. quantum physics}
Quantum physics and thermodynamics are theories of randomness. Therefore, they have deeply influenced each other, both in the understanding of randomness and the program such understanding leads to. Here we analyze some of these connections for two branches of thermodynamics, i.e. statistical physics and stochastic thermodynamics. We conclude by new potential applications of quantum randomness opened in the emerging field of quantum thermodynamics.

\subsubsection*{Statistical physics}

Since it was the first physical theory dealing with randomness, statistical physics has inspired the earliest views on quantum randomness. 
Within the framework of classical statistical physics, randomness is purely epistemic: The "real" states are the micro-states pertaining to elementary systems, whose evolution is deterministic and reversible. The micro-states are hidden at the macroscopic level, because of the coarse-graining operated by our perception and instruments. Randomness captures the effectiveness of our description at this scale.\\

\noindent From a thermodynamical point of view, this coarse-graining is the very origin of the Second Law, which historically states that "The entropy of an isolated system can only increase". Actually, such strictly positive entropy production corresponds to the amount of information lost while zooming out from the microscopic to the macroscopic world. All these arguments have contributed to build a hierarchy in the physical theories, classical
thermodynamics appearing as some effective theory where randomness and irreversibility are
weaknesses due to our lack of knowledge on physical states. On the contrary,
the ultimate description of physical reality is reached at the microscopic level where randomness
vanishes and the physical laws are reversible. It is the program of reductionist theories such as statistical physics to describe
the physical phenomena at such ultimate level.\\

\noindent As mentioned in the introduction, quantum randomness was originally conceived as some kind of irreducible epistemic randomness. Applying the program of statistical physics has originally lead to the search for hidden variables (equivalent to the micro-states) and attempts to get rid of the measurement postulate (equivalent to the imperfect coarse graining step). Even nowadays, after local hidden variable theories have been discarded by the multiple violations of Bell's inequalities, the initial program of statistical physics keeps inspiring some interpretations currently defended, e.g. decoherence theories and their subsequent approaches based on quantum darwinism \cite{WZ1,WZ2} where the spirit of reductionism is still vivid. In these views, the whole world is made of elementary systems whose states (the wave functions) are hidden and whose evolution is governed by the deterministic Schr\"odinger equation. Just like in statistical physics, here the program is to let the classical world emerge from larger and larger samples of quantum systems - but lets open the question of the origin of the quantum formalism, which is postulated \cite{PT}.  \\

\begin{figure}[!h]
\centering\includegraphics[scale=0.35]{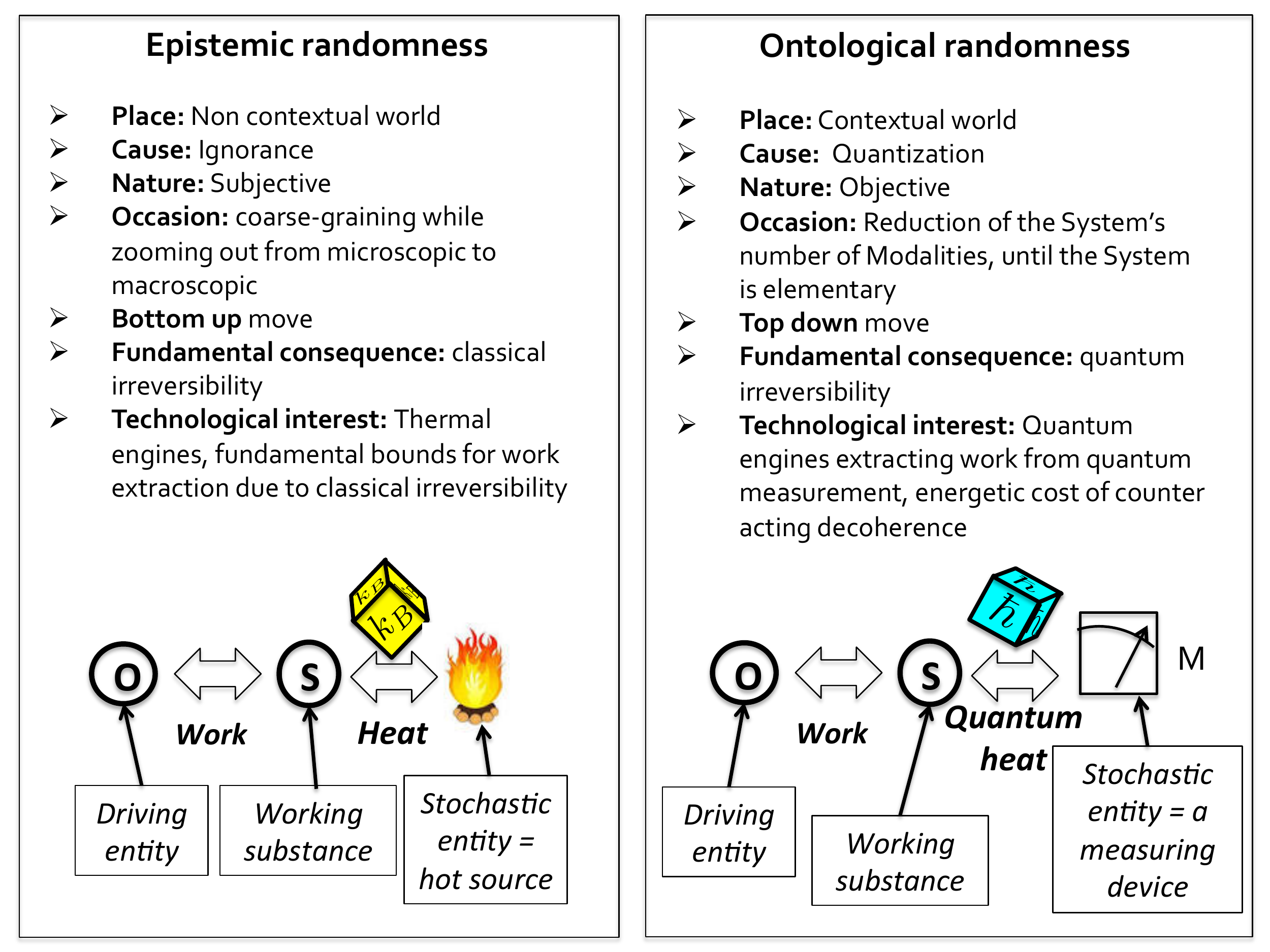}
\caption{The two dices of physics. Left: "$k_B$ dice". Epistemic randomness is caused by ignorance and coarse graining. It has given birth to classical thermodynamics and classical heat engines, which extract work from the classical noise induced by classical randomness. Right: "$\hbar$ dice". Ontological randomness is caused by contextuality and quantization. It gives rise to a new framework for quantum thermodynamics, where the stochastic entity is the context/measuring device. New engines of genuinely quantum essence can be designed, that allow for work extraction from the quantum noise induced by quantum randomness only. }
\label{fig_thermo}
\end{figure}

\subsubsection*{Stochastic thermodynamics}
Since the nineties, stochastic thermodynamics aims to extend the Laws of thermodynamics to small, possibly quantum physical systems driven out of equilibrium by some external operator \cite{Seifert}. In addition to the driving entity, these systems interact with a reservoir, that involve a large number of other systems. This interaction is described in an effective manner, by a stochastic term in the equation describing the system's evolution. Under the action of the drive and the reservoir, the system describes a stochastic trajectory in its phase space, consisting of sequences of continuous evolution, interrupted by stochastic jumps.A major progress brought by stochastic thermodynamics has been to unveil the central role of randomness in the understanding of (i) the nature of heat engines and (ii) the thermodynamic time arrow:

\begin{enumerate}
\item Firstly, randomness allows defining a clear demarcation line between two different kinds of energy exchanges: Heat and work. Work is exchanged between the system and the driving entity and corresponds to organized, useful, re-usable energy. Heat is randomly exchanged between the system and reservoir during the jumps. It is the purpose of heat engines to turn heat into work - in other words, to extract useful energy from random noise. Random noise is the thermodynamical resource of heat engines.\\

\item Secondly, randomness is the cause of irreversibility at the scale of single trajectories - where the laws of physics are in principle reversible. The idea relies on the simple Gedanken experiment: (i) drive the system between time $t=0$ and $t=T$. (ii) Invert the protocol to bring the system back to its initial state. The absence of random perturbation signals full control on the system, which is prepared back to its initial state. This corresponds to a perfectly reversible transformation. On the contrary, any random perturbation can prevent the transformation from being reversible. The second Law can be recovered in this new framework, which provides efficient tools to quantify irreversibility by the amount of entropy production. 
\end{enumerate}

\subsubsection*{Outlook: Rebuilding thermodynamics on quantum randomness}
Stochastic thermodynamics has shed new light on the basic purposes of thermodynamics : How to extract the maximal amount of work from a stochastic energy source? How is work extraction constrained by the Second Law - in other words, what are the fundamental bounds for work extraction? Reciprocally, what is the energetic cost to perform a given protocol, in the presence of random noise? In some sense, thermodynamics could be recalled {\it the art of noise}. \\

\noindent Interestingly, the very causes of randomness in stochastic thermodynamics do not matter. It was originally developed to deal with epistemic randomness, for it involved reservoirs with large numbers of degrees of freedom whose complete description is practically impossible. However, the concepts and methods of stochastic thermodynamics solely rely on the notion of stochastic trajectory. In principle, they can thus be applied to whichever source of randomness. This motivated some of us to adapt them to the case where randomness is of quantum origin and induced by a change of context - in the standard terminology of quantum mechanics, by quantum measurement itself \cite{QMES1,QMES2,QMES3} (Fig.2). \\

\noindent This new framework has lead to identify a genuinely quantum component to heat, corresponding to the energetic fluctuations induced by quantum measurement. Such energetic fluctuations called quantum heat are the fuel of a new type of quantum engine, which can produce work in the absence of any heat source, but just because the working substance is measured \cite{QMES2}. Consequently, quantum randomness can be seen as a genuinely quantum resource, whose origin is measurement induced quantum back-action on quantum systems. These new generations of quantum devices offer an appealing alternative application of quantum randomness, in addition to random number generation. Reciprocally, this new framework opens avenues to investigate the energetic cost of counter-acting quantum noise. \\

\noindent In the same way, quantum randomness provides a genuinely quantum cause for irreversibility, that is now induced by stochastic quantum measurements. It is an open line of research to investigate how such quantum entropy production impacts the novel quantum mechanisms of work extraction, as well as the energetic costs of quantum control \cite{viewpoint}.  
This new facet of thermodynamics thus provides the adapted framework to assess the energetic sustainability of quantum technologies, for which the biggest enemy has for long been identified as decoherence.

\section{Conclusion}
Sources of randomness known in the classical world are either intrinsic and metaphysically postulated, or epistemic. Epistemic randomness is caused by the ignorance of a given system's state and grows in the non-contextual world we are accustomed to. Cornerstone of classical thermodynamics, epistemic randomness has lead to develop powerful concepts of both practical and fundamental interest, such as the thermodynamic arrow of time and optimal work extraction in heat engines.\\

\noindent In this article we have considered the case of a physical world characterized by contextual states called modalities. A new kind of randomness of ontological nature is shown to appear if the number of exclusive modalities is discrete, which is the property of elementary systems. Ontological and quantum randomness are the consequence of the fundamental mismatch between the quantized nature of such exclusive modalities, and the continuous nature of the contexts' state. In this approach called CSM, the quantum formalism is not meant to describe absolute properties of systems' evolution, but deals with joint properties of systems and contexts. CSM is thus not compatible with classical reductionism, which pre-supposes the non-contextuality of states.

Since it provides a clear ontological basis for quantum randomness, CSM has lead to recent propositions aiming to rebuild thermodynamics in a new landscape. While classical heat engines extract work from classical noise, new generations of quantum engines have been proposed, that are fueled by quantum noise. These devices constitute new technological applications of quantum randomness. \\

\noindent At the crossing of physical and philosophical roads, CSM sheds new light on the relations between epistemology, objectivity and ontology. These relations have for long been established, in a non-contextual world allowing for absolute objectivity. They should now be totally revisited in the new scenery opened by contextual objectivity.\\

{\it The authors thank Nicolas Gisin, \u{C}aslav Bruckner and Borivoce Daki\'c for challenging and enlightening discussions, as well as Nayla Farouki for her constant and invaluable support.}



\end{document}